# Entropy of leukemia on multidimensional morphological and molecular landscapes

Jose M. G. Vilar[1,2]

[1]*Biophysics Unit (CSIC-UPV/EHU) and Department of Biochemistry and Molecular Biology, University of the Basque Country, P.O. Box 644, 48080 Bilbao, Spain*

[2]*IKERBASQUE, Basque Foundation for Science, 48011 Bilbao, Spain*

## Abstract

Leukemia epitomizes the class of highly complex diseases that new technologies aim to tackle by using large sets of single-cell level information. Achieving such goal depends critically not only on experimental techniques but also on approaches to interpret the data. A most pressing issue is to identify the salient quantitative features of the disease from the resulting massive amounts of information. Here, I show that the entropies of cell-population distributions on specific multidimensional molecular and morphological landscapes provide a set of measures for the precise characterization of normal and pathological states, such as those corresponding to healthy individuals and acute myeloid leukemia (AML) patients. I provide a systematic procedure to identify the specific landscapes and illustrate how, applied to cell samples from peripheral blood and bone marrow aspirates, this characterization accurately diagnoses AML from just flow cytometry data. The methodology can generally be applied to other types of cell-populations and establishes a straightforward link between the traditional statistical thermodynamics methodology and biomedical applications.

# I. INTRODUCTION

Many complex diseases require the measurement of multiple molecular factors at the single-cell level over large populations of cells for their precise characterization and detailed understanding [1-3]. Current technologies, such as flow cytometry (FCM), allow for such measurements, including simultaneous quantification of several morphological and molecular properties [4,5]. These technologies can currently provide the simultaneous single-cell measurement of tens of surface and intracellular markers of up to thousands of cells per second [6,7]. Such rapid technological development, however, is yet to be matched by traditional analysis tools to interpret the data [8,9]. It is now clear that all this data by itself, without the analytical tools to extract the relevant information, is not enough to faithfully understand the underlying cellular processes and their dysregulation in diseases such as cancer.

A prototypical example in which large amounts of data are generated is the cytometric analysis of acute myeloid leukemia (AML), a type of cancer produced by the dysregulated growth of the myeloid line of blood cells [10]. AML leads to abnormal white blood cells, red blood cells, or platelets to accumulate in the bone marrow and blood, which interferes with the production of normal blood cells. The presence of abnormal cells can potentially be detected by analyzing the changes of the distribution of morphological and molecular attributes of cell samples from blood or bone marrow. There are important difficulties associated with this approach for diagnosing AML. Besides abnormal cells being mixed with normal cells in the population samples, there are several AML subtypes and there is not an obvious well-defined set of changes in the molecular attributes that can fully characterize AML. In addition, there is the natural variation between individuals that overlaps with the changes induced by AML. Thus, the main challenge is to identify the quantities that best can be used to distinguish between AML patients and Normal (AML-free) individuals based on the single-cell statistical properties of their blood or bone marrow cell populations.

Here, I show that entropy, as traditionally used in statistical thermodynamics [11], provides a measure for the precise diagnosis of AML. Cell populations are characterized by their entropies in multidimensional landscapes constructed from the distributions of single-cell morphological and molecular attributes of flow cytometry data. Diagnosis is achieved by comparing how far the cell population distribution of an individual is from the AML and Normal prototypical maximum-entropy distributions.

The approach presented here was applied to samples from peripheral blood and bone marrow of 359 patients provided at the DREAM6/FlowCAP2 Molecular Classification of AML Challenge [9,12]. The results accurately diagnosed AML in a blind set of 179 patients from just the provided training set of 180 patients, and ranked first among the best performing methods of this challenge [9,13]. Challenges such as those put forth by DREAM (Dialogue for Reverse Engineering Assessments and Methods) and FlowCAP (Flow Cytometry: Critical Assessment of Population Identification Methods) provide objective unbiased evaluations of computational methods in complex situations and eliminate the possibility of overfitting since the results are not known in advance [9,14-16]. The different approaches that participated at the DREAM6/FlowCAP2 challenge included mostly traditional machine learning algorithms, such as Support Vector Machine Regression, Logistic Regression, Vector Quantization, and Correlative Matrix Mapping. Overall, most of the approaches obtained good results [9].

## II. GENERAL APPROACH

Leukemia, as any other type of cancer, results from dysregulated growth caused by genetic and epigenetic changes that alter the cellular state [17]. Therefore, the first step towards distinguishing cancerous from normal cell populations is to characterize the state of each cell. To take into account that there is only limited information, the approach considers measured, $x$, and internal (non-measured), $q$, quantities separately. In the case of FCM, light scattering and the intensity of fluorescent reporters are the prototypical examples of measured quantities. Internal quantities are much more numerous and include, among others, non-measured protein levels and specific DNA mutations. The characterization of a population $i$ considers the probability distribution $P_i(x,q)$ of these two types of attributes among its cells.

The goal is to discriminate among different cell-population types based on the statistical properties of their measurable quantities. To be able to take into account the effects of the internal quantities $q$, one must estimate their effects from the measurable quantities $x$, which should carry sufficient information about the key cellular differences between the different population types.

A convenient starting point to estimate the effects of the internal quantities in a cell population is the entropy of its attributes distribution $S_i = -\int P_i(x,q) \ln P_i(x,q) dxdq$. This quantity can be expressed as a function of just $x$ by rewriting the join probability $P_i(x,q) = P_i(q \mid x) P_i(x)$ in terms of the conditional probability of $q$ with respect to $x$, $P_i(q \mid x)$, and the probability of $x$, $P_i(x)$. Integration over $q$ leads to

$$S_i = -\int P_i(x)(f_i(x) + \ln P_i(x))dx\,, \qquad (1)$$

which explicitly encapsulates the contributions from internal quantities into the function $f_i(x) = \int P_i(q \mid x) \ln P_i(q \mid x) dq$.

The main hypothesis to proceed further is that the internal contribution for each population $f_i(x)$ can accurately be approximated by the same function $f_T(x)$ for all the members of a given type $T$. In the case of AML and Normal types, the result would be $f_i(x) = f_{AML}(x)$ or $f_i(x) = f_{Normal}(x)$ depending on whether the population $i$ is AML or Normal.

The function $f_T(x)$ can be estimated by considering a reference maximum entropy distribution $P_T(x)$ that integrates the main features of the cell populations of the type $T$. The maximum condition implies that $\delta S_T = -\int \delta P_T(x)(f_T(x) + \ln P_T(x))dx$ is zero, which happens when the term multiplying the variation of the probability in the integrand is constant. This condition can be rewritten as $f_T(x) = -\ln P_T(x) - S_T$, where the constant $S_T$ is the maximum entropy for the cell population type $T$. Therefore, using $f_i(x) = f_T(x) = -\ln P_T(x) - S_T$ in Eq. (1) leads to

$$S_{i,T} = S_T - \int P_i(x) \ln(P_i(x) / P_T(x))dx\,, \qquad (2)$$

which can be interpreted as the entropy of the population $i$ on the multidimensional morphological and molecular landscape defined by the maximum entropy distribution for the cell population type $T$.

The explicit reference distribution $P_T(x)$ is chosen as the distribution that maximizes the total entropy of all the members of the type $T$. The variation of the total entropy with respect to $P_T(x)$ is given by $\delta(\sum_{i \in T} S_{i,T}) = -\int \delta P_T(x)(\sum_{i \in T} P_i(x) / P_T(x))dx$, which is zero when $P_T(x)$ is proportional to $\sum_{i \in T} P_i(x)$. Therefore, $P_T(x)$ is chosen to be the average distribution, which for AML and Normal types is given by $P_{AML}(x) = (1/N_{AML})\sum_{i \in AML} P_i(x)$ and $P_{Normal}(x) = (1/N_{Normal})\sum_{i \in Normal} P_i(x)$, respectively, where $N_{AML}$ and $N_{Normal}$ are the number of cell populations for each type.

The explicit form of Eq. (2) shares many similarities with other expressions used in physical sciences and in information theory. It has exactly the same form as the Gibbs entropy formula [11,18,19]. The main difference is that, in that case, the state $T$ refers to an equilibrium state instead of a cell population type. It is also very similar, except for the constant $S_T$, to the Kullback-Leibler divergence, which uses relative entropies to compare distributions [20]. It is important to emphasize that entropy defined by Eq. (2) does not quantify variability. Instead, it quantifies how similar the distribution of measured values for a given individual is to the expected average distribution for a given type. High entropy means that the distribution of the individual is similar to the expected distribution and low entropy means that it is different. Both high and low variability would lead to low relative entropy values if the underlying distributions are not close to the expected distribution.

A key feature of the method is that since it is based on a maximum principle with respect to the distributions of the attributes, the effects of small perturbations in the measured distributions are second order and therefore they will only impact minimally the results.

The analogy with statistical thermodynamics can be extended further to estimate the likelihood $p_{i,T}$ of the population's belonging to the type $T$ through the Einstein fluctuation formula, $p_{i,T} \sim e^{S_{i,T}}$, which relates the probability of observing a state with its entropy [21,22]. After normalization, it results in

$$p_{i,T} = \frac{e^{S_{i,T}}}{\sum_{T'} e^{S_{i,T'}}}, \qquad (3)$$

where the sum in the denominator is performed over all the cell population types. This expression assigns a high likelihood to a cell population as being of type $T$ if the distribution of its attributes has a form similar to the maximum entropy distribution for that type. This assignment parallels in many ways the approach first used by A. Einstein to compute the probability of observing a fluctuation moving a system away from its equilibrium state based on its entropy change [22].

## III. APPLICATION TO DIAGNOSING LEUKEMIA

The explicit application to diagnosing AML relies on evaluating whether the distribution of the values of flow cytometry data for a given individual to be diagnosed (Fig. 1 and Supplementary Fig. 1) is closer to the AML or Normal maximum entropy distribution (Fig. 1). Using just a few variables, such as side scatter and a fluorescent marker for the receptor protein CD45, is informative in many cases but does not offer a full characterization (compare for instance AML patient #7 with Normal patient #96 in Supplementary Fig. 1). In general, there are several sets of simultaneously measured variables for a given individual and therefore there are several different landscapes. For each population $i$ and each set $k$ of simultaneously measured variables $x^k$, the corresponding entropy is given by $S_{i,T}^k = S_T^k - \int P_i\,(x^k)\ln(P_i\,(x^k)/P_T(x^k))dx^k$. For an AML individual, the values of the entropies are expected to be $S_{i,AML}^k - S_{AML}^k \approx 0$ and $S_{i,Normal}^k - S_{Normal}^k << 0$. A Normal individual, in contrast, would lead to $S_{i,AML}^k - S_{AML}^k << 0$ and $S_{i,Normal}^k - S_{Normal}^k \approx 0$. Therefore, the relative entropy difference

$$\Delta S_i^k = S_{i,Normal}^k - S_{i,AML}^k = S_{Normal}^k - S_{AML}^k - \int P_i\,(x^k)\ln(P_{AML}(x^k)/P_{Normal}(x^k))dx^k\ , \qquad (4)$$

indicates that the individual looks like an AML patient for negative values and like a Normal subject for positive values. These entropies can be used to compute the weighted entropy difference

$$\Delta\tilde{S}_i = \sum\nolimits_k w_k \Delta S_i^k\ , \qquad (5)$$

where $w_k$ is the weight of the set $k$. Similarly, the weighted entropies are defined by $\tilde{S}_{i,T} = \sum_k w_k S_{i,T}^k$ and $\tilde{S}_T = \sum_k w_k S_T^k$. The likelihood for the population $i$ to be AML is estimated using Eq. (3) with the weighted entropy difference as

$$p_{i,AML} = \frac{1}{1+e^{\Delta\tilde{S}_i}}\ . \qquad (6)$$

## IV. IMPLEMENTATION

The DREAM6/FlowCAP2 data [9,12] consists of seven groups of measurements for each individual, each group comprising the simultaneous measurements of the forward scatter (FS), the side scatter (SS), a fluorescent marker for the receptor protein CD45 (FL3), and a different combination of four different fluorescent markers for other proteins (FL1, FL2, FL4, and FL5). The specific fluorescent markers used as FL1, FL2, FL4, and FL5 in each group of measurements are listed in Table I. Therefore, there are seven different attribute spaces, described by the seven-dimensional vectors $x =$(FS, SS, FL1, FL2, FL3, FL4, FL5). For each space, the entropy can be computed in the full seven dimensions or in any combination of subspaces.

An avenue to estimate the suitability of a space or subspace for AML and Normal individual discrimination is to use leave-one-out cross-validation with the training data. In this case, it consists in testing each individual $i$ of the training set without its contribution to $P_{AML}$ or $P_{Normal}$.

The results of cross-validation with uniform weights $w_k = 1$ (Fig. 2 and Supplementary Fig. 2) indicate that Normal and AML populations have high and low $\Delta\tilde{S}_i$ values respectively, as expected from the assumptions of the approach. Only the population samples of just a few AML individuals have higher $\Delta\tilde{S}_i$ than the lowest $\Delta\tilde{S}_i$ for the Normal set. Therefore, the assumption that the function $f_i(x)$ can be approximated by the maximum entropy distribution of its type holds for the cell populations of most individuals. In general, it is observed that for higher dimensionality spaces, the number of AML individuals that overlap with Normal ones is smaller. Such a small overlap indicates that the entropy on multidimensional morphological and molecular landscapes provides an efficient avenue to discriminate AML from Normal populations.

The results with the test set, using $P_{AML}$ and $P_{Normal}$ computed from the training set, show even better segregation of individuals based on the entropy of their cell-population samples (Fig. 2 and Supplementary Fig. 2). It is possible to use Eq. (6) to estimate the likelihood for an individual of being AML positive. However, the values $S_{AML}$ and $S_{Normal}$ for the maximum entropy of AML and Normal individuals are not readily available. A first approximation, based on the cross-validation results, is to consider $S_{AML} - S_{Normal} \approx 0$. The results indicate that the approach accurately diagnoses AML, except for just a few individuals (Fig. 2 and Supplementary Fig. 2) and that the performance increases with the dimensionality of the space in which the entropy is computed. The results also suggest that the test data was easier to classify than the training data, which contains a few difficult patients.

The discriminative capabilities can be improved by selecting, among the many possibilities available, not only the dimensionality of the space but also the subspaces. The subspace can be chosen by selecting the weights $w_k$ to be either 0 or 1. For instance, a combination that proved to give good results is the 4-dimensional distributions for the values of $x^k =$ (FS, SS, FL3, $k$) with $k \in \{$FL1, FL2, FL4, FL5$\}$ for the seven groups of measurements. Taking into account that the markers denoted as FL1, FL2, FL4, and FL5 are different in each group of measurements, there are $7 \times 4 = 28$ landscapes for each individual. The predictions obtained with this approach ranked first among the best performers of the DREAM6 competition (Supplementary Information). Further improvement can be achieved by adjusting the values of the weights to maximize the segregation between AML and Normal in the cross-validation phase.

A systematic procedure for adjusting the weights $w_k$ in Eq. (5) to better segregate between AML and Normal individuals is to increase the weights that benefit the segregation and decrease the weights that worsen it. Segregation is quantified by the distance between the Normal, $N_{\min}$, and AML, $A_{\max}$, individual with the minimum and maximum entropy difference, respectively. This distance is explicitly defined by $\Delta\Delta\tilde{S} = \sum_k w_k \Delta\Delta S^k$, with $\Delta\Delta S^k = \Delta S^k_{i=N\min} - \Delta S^k_{i=A\max}$. The procedure considers updates proportional to a small quantity $\Delta t$ so that the weights at the step index $t$ are updated iteratively to the step index $t + \Delta t$ through the expression

$$w_k(t+\Delta t) = \frac{w_k(t)(1+\Delta t \Delta\Delta S^k)}{\sum_k w_k(t)(1+\Delta t \Delta\Delta S^k)}. \quad (7)$$

This procedure guaranties that $\Delta\Delta\tilde{S}$ increases in each interaction if $\Delta t$ is sufficiently small and the identities of the individuals $N_{\min}$ and $A_{\max}$ do not change. Under these assumptions, expanding Eq. (7) in $\Delta t$ and using $\Delta\Delta S(t)^2 = \sum_k \Delta\Delta S(t)^2 w_k(t)$, leads to $\Delta\Delta\tilde{S}(t+\Delta t) \approx \Delta\Delta\tilde{S}(t) + \Delta t \sum_k (\Delta\Delta S^k - \Delta\Delta S(t))^2 w_k(t)$, which shows that the increase is proportional to a combination of positive quantities.

The enhancement of the discrimination between AML and Normal individuals by adjusting the weights $w_k(t)$ according to Eq. (7) is illustrated by the values of the entropies of each population $i$ with respect to AML, $\tilde{S}_{i,AML} - \tilde{S}_{AML}$, and Normal, $\tilde{S}_{i,Normal} - \tilde{S}_{Normal}$, states. For $w_k = 1$, AML and Normal populations tend to be separated according to their entropies but there is a significant overlap (Fig. 3a). By updating the weights $w_k$ according to Eq. (7), the quantity $\Delta\Delta\tilde{S}$, which is negative, is increased by decreasing its absolute value. This quantity never changes to positive values and there is not an actual segregation of the populations (Fig. 3b). Closer inspection indicates that segregation is prevented by a patient in the AML training set (patient #116 in Fig. 1). By removing this patient from the calculation of $A_{\max}$, there is a clear segregation of AML and Normal individuals according to the entropies of their distributions (Fig. 3c). Perfect segregation occurs in both cross-validation with the training set, which was used to obtain the different parameters, and predictions with the test set, which is completely independent of the training set. As shown in Fig. 3c, the reason patient #116 prevented segregation is because this patient has the Normal instead of the typical AML signatures.

The segregation measure $\Delta\Delta\tilde{S}(t)$ starting with $w_k = 1$ increases with the number of iterations, as shown mathematically, until it reaches a plateau. In all the cases except for 7-D, it changes its sign from negative to positive values (Supplementary Fig. 3), implying that complete segregation has been achieved in the training set. This segregation, except for 7-D, is also present in the test set (Fig. 4 and Supplementary Fig. 4). In the 7-D case, $\Delta\Delta\tilde{S}(t)$ remains negative, indicating that such a high dimensional space is not fully reliable for segregation based on entropies. By choosing $\tilde{S}_{AML} - \tilde{S}_{Normal}$ so that $\sum_k w_k(t_{\max})(\Delta S^k_{i=N\min} + \Delta S^k_{i=A\max})/2 = 0$, the approach correctly diagnoses the presence of AML in all the cases of the test set from one- to six- dimensions (Fig. 4 and Supplementary Fig. 4).

## V. DISCUSSION

The use of physical approaches has been remarkably successful in the characterization of heterogeneous cell populations and their evolution in many complex biological scenarios [23-26] to the extent of making inroads in the mainstream biomedical research [27,28]. One of the main problems posed by hematological cancers, such as AML, is the underlying heterogeneity resulting from diverse molecular changes in the cellular state, including several recurrent mutations and chromosome translocations [10,17]. This heterogeneity is responsible to a large degree for the observed different clinical courses and is characterized at the cellular level by changes in size and granularity and by the acquisition and loss of characteristic cell surface markers. Traditional approaches to diagnose AML seek to identify subpopulations of cells with these characteristic changes induced by the disease [2,9].

The analysis presented here has shown that, despite the heterogeneity present, it is possible to define an average characterization of AML and Normal cell populations on multidimensional morphological and molecular landscapes that can be used to accurately diagnose the presence of this type of cancer. This approach shares many similarities with traditional statistical thermodynamic methods that use entropy to collapse many nonmeasurable degrees of freedom within just a few key macroscopic or mesoscopic variables. In the case of cell populations, the quantities analogous to the macroscopic and mesoscopic variables are the combinations of attributes that define multidimensional morphological and molecular landscapes for each population type. This analogy establishes a straightforward link between statistical thermodynamics and biomedical applications and opens the door to the use of the sophisticated statistical physics methodology to tackle problems of direct medical importance.

## ACKNOWLEDGMENTS

I thank Wade T. Rogers for providing the data for the DREAM6/FlowCAP2 Molecular Classification of Acute Myeloid Leukemia Challenge prior to publication and for stimulating discussions. This work was supported by the MINECO under grant FIS2012-38105.

---

## TABLES

TABLE I. List of the five specific fluorescent markers, denoted by FL1, FL2, FL3, FL4, and FL5, used in each of the 7 groups of measurements.

| Group | FL1 | FL2 | FL3 | FL4 | FL5 |
|---|---|---|---|---|---|
| **1** | IgG1-FITC | IgG1-PE | CD45-ECD | IgG1-PC5 | IgG1-PC7 |
| **2** | Kappa-FIT | Lambda-PE | CD45-ECD | CD19-PC5 | CD20-PC7 |
| **3** | CD7-FITC | CD4-PE | CD45-ECD | CD8-PC5 | CD2-PC7 |
| **4** | CD15-FITC | CD13-PE | CD45-ECD | CD16-PC5 | CD56-PC7 |
| **5** | CD14-FITC | CD11c-PE | CD45-ECD | CD64-PC5 | CD33-PC7 |
| **6** | HLA-DR-FITC | CD117-PE | CD45-ECD | CD34-PC5 | CD38-PC7 |
| **7** | CD5-FITC | CD19-PE | CD45-ECD | CD3-PC5 | CD10-PC7 |

## FIGURE CAPTIONS

FIG. 1 (color online). Molecular and morphological landscapes of leukemia. The two-dimensional distribution of cells with given values of the logarithms of the side scatter (SS log) and the marker CD45 (CD45 log) intensities are shown for representative cell populations, with black and white colors representing high and zero densities, respectively. The distributions for AML patient #26 ($P_{26}$) and normal individual #4 ($P_4$) are indicative of distributions that closely resemble the maximum entropy distribution of their state, either AML ($P_{AML}$) or Normal ($P_{Normal}$) states. The distribution for AML patient #116 ($P_{116}$), in contrast, does not show a clear discriminative feature of the presence of AML in the SS-CD45 landscape.

FIG. 2 (color online). Entropy-based discrimination between AML and Normal individuals. The total relative entropy difference, using Eq. (5) with $w_k = 1$, is shown for each individual of the training and test sets for two-, three-, and four-dimensional landscapes. The likelihood that an individual of the test set is AML positive, given by Eq. (6), is also shown for two-, three-, and four-dimensional landscapes. Circles and squares represent, respectively, the actual AML and Normal state of a patient as clinically assessed by a physician. The insets are magnifications around the AML-Normal boundary.

FIG. 3 (color online). Total and weighted entropies on the two-dimensional landscapes defined by the maximum entropy distribution for AML and Normal states. The entropies for the training (left column) and test (right column) sets were computed from Eq. (5) with uniform $w_k$ (**a**), with the $w_k$ that maximize the distance $\Delta\Delta\tilde{S}$ of the training set (**b**), and with the $w_k$ that maximize the distance $\Delta\Delta\tilde{S}$ of the training set excluding patient #116 (**c**). Circles and squares represent, respectively, the actual AML and Normal state of a patient as clinically assessed by a physician. The dashed lines are given by $\tilde{S}_{Normal} - \tilde{S}_{i,Normal} = \tilde{S}_{AML} - \tilde{S}_{i,AML} + \Delta\Delta\tilde{S}(t_{\max})/2$ with the value of $\Delta\Delta\tilde{S}(t_{\max})/2$ equal to 0 (**a**), -5.21 (**b**), and 8.15 (**c**). The black circle with the orange contour corresponds to the AML patient #116.

FIG. 4 (color online). Weighted-entropy enhancement of the discrimination between AML and Normal individuals. Same situation as in Fig. 2 but with the weights $w_k$ that maximize the distance $\Delta\Delta\tilde{S}$ of the training set excluding patient #116.

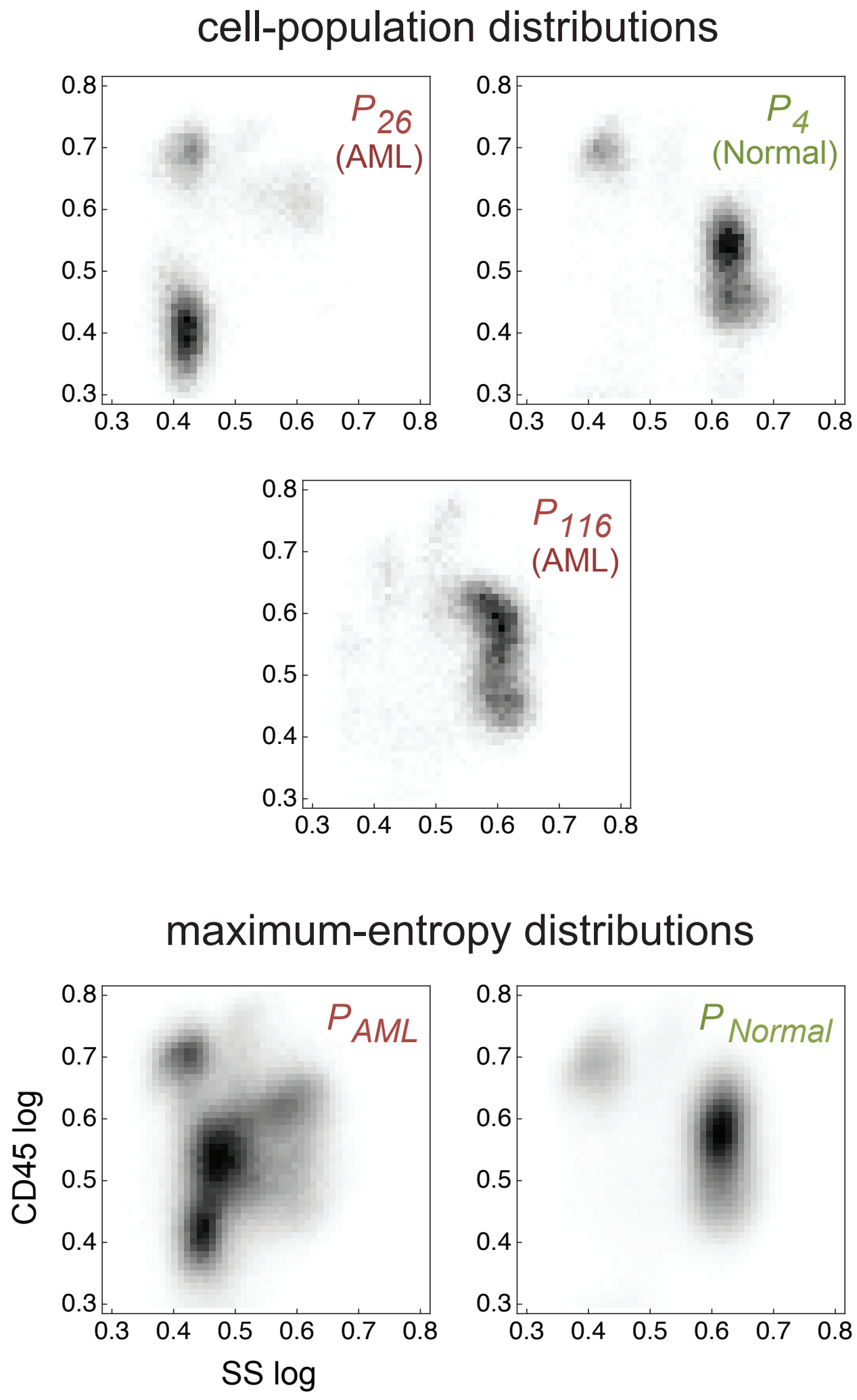

Figure 1

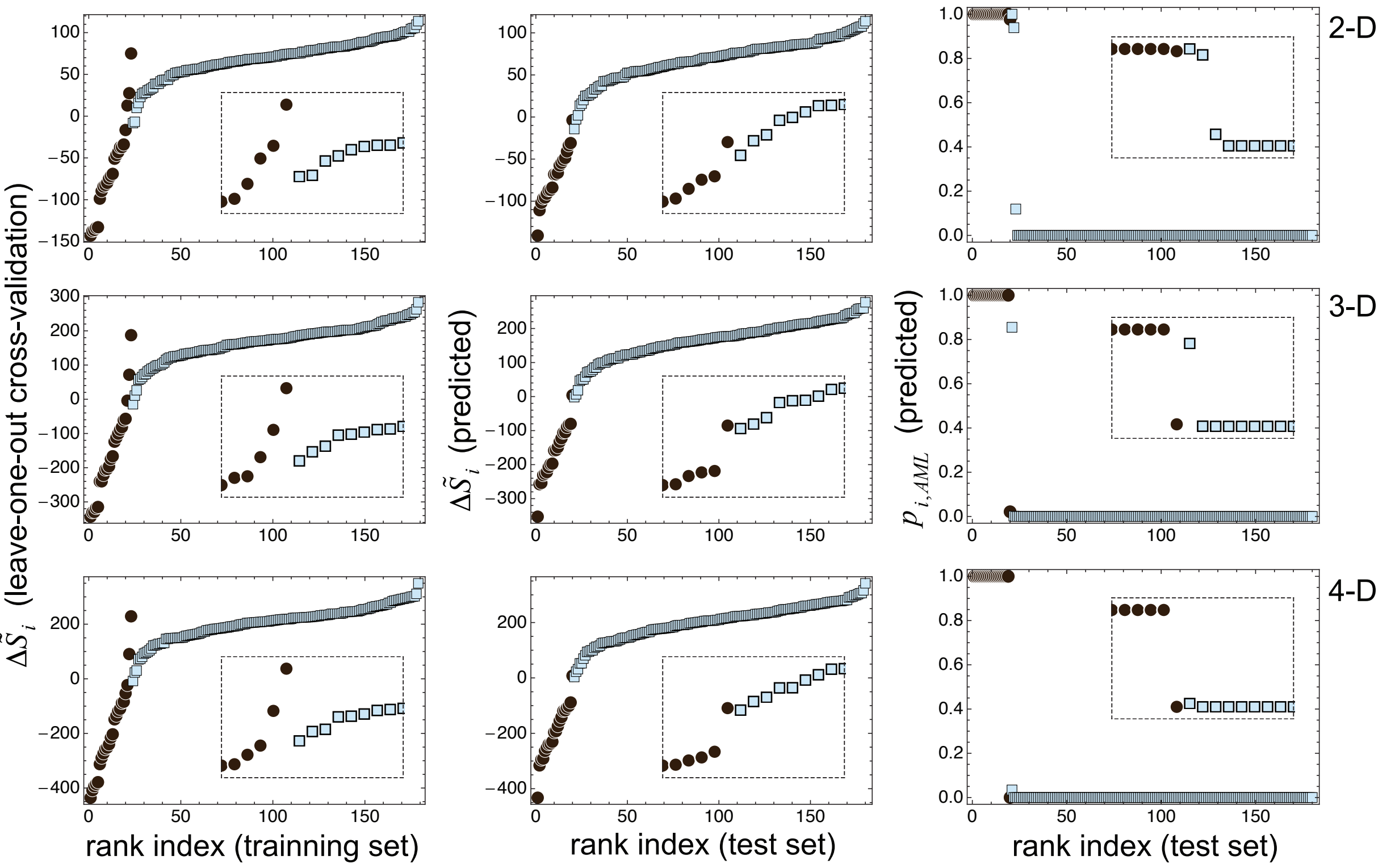

Figure 2

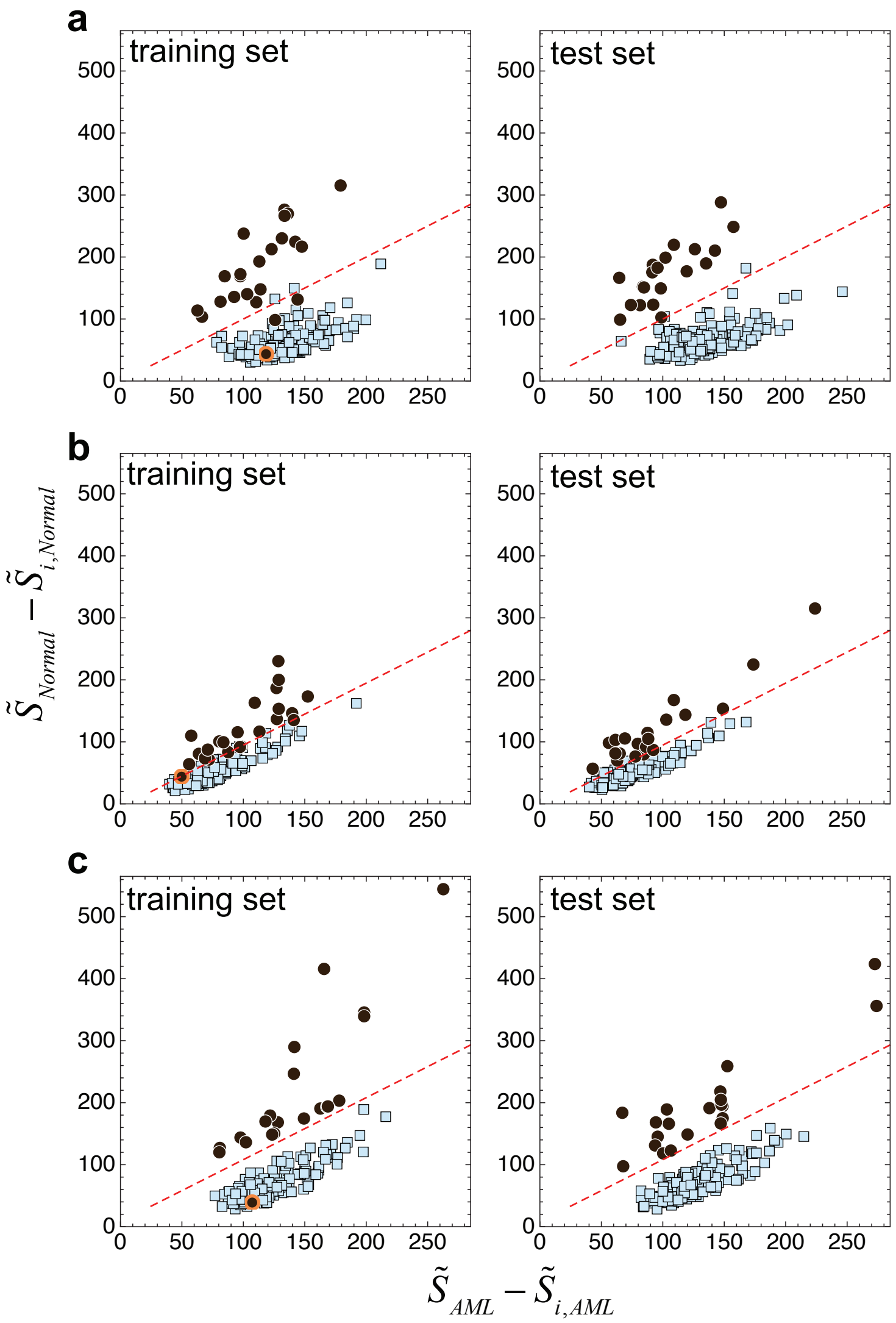

Figure 3

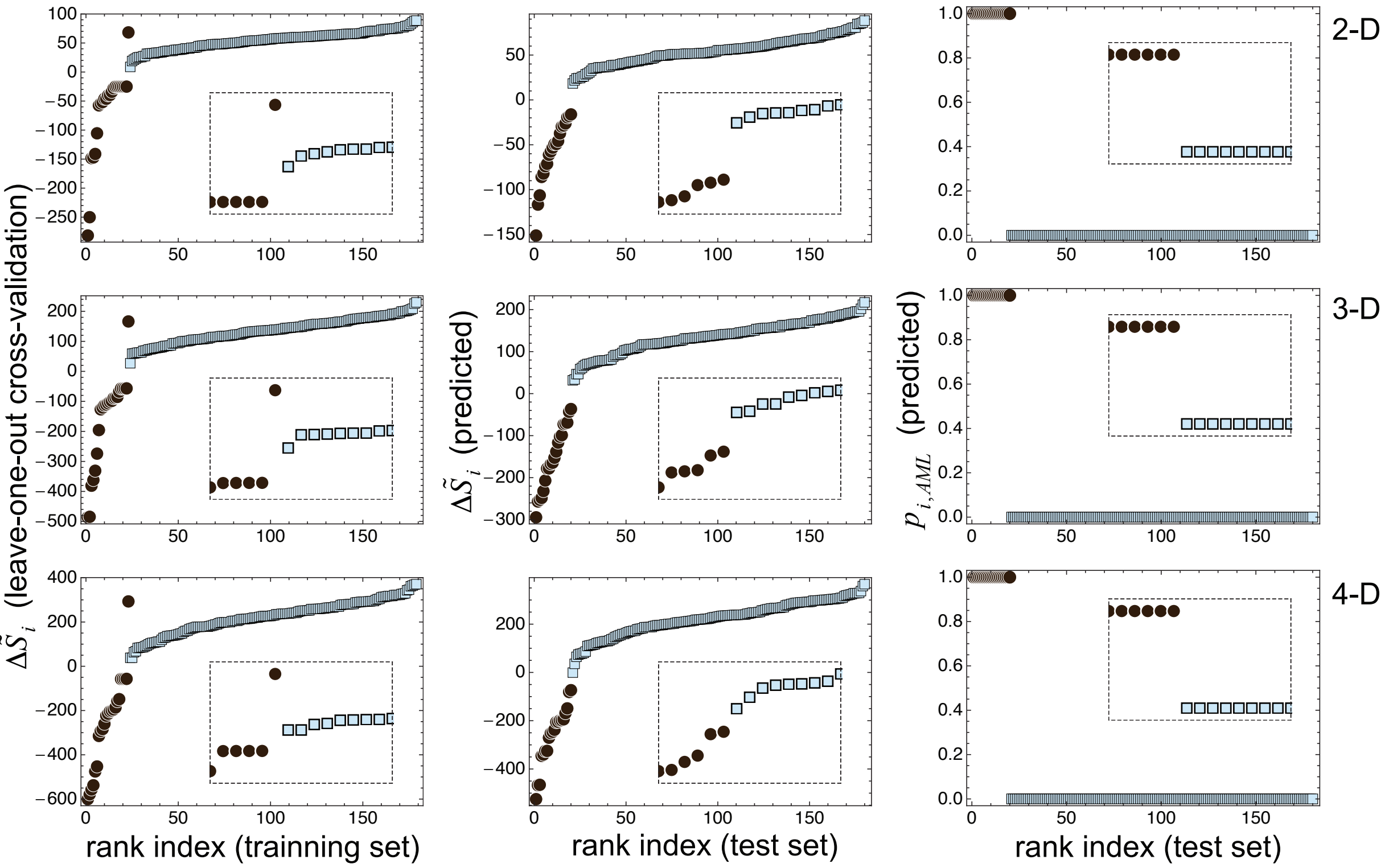

2-D
3-D
4-D
$\Delta\tilde{S}_i$ (leave-one-out cross-validation)
$\Delta\tilde{S}_i$ (predicted)
$p_{i,AML}$ (predicted)
rank index (trainning set)
rank index (test set)
rank index (test set)

Figure 4

# Supplementary Information: Entropy of leukemia on multidimensional morphological and molecular landscapes

*Jose M. G. Vilar*[†‡]
[†]*Biophysics Unit (CSIC-UPV/EHU) and Department of Biochemistry and Molecular Biology, University of the Basque Country, Bilbao, Spain*
[‡]*IKERBASQUE, Basque Foundation for Science, Bilbao, Spain*

This Supplementary Information includes the details of the implementation of the Team21's solution to the DREAM6/FlowCAP2 Molecular Classification of Acute Myeloid Leukemia Challenge; the source code of two python files; and four supplementary figures.

## Team21's solution to the DREAM6/FlowCAP2 Molecular Classification of Acute Myeloid Leukemia Challenge

The approach of *team21*, consisting of *Jose M. G. Vilar*, ranked first (http://www.the-dream-project.org/result/classification-aml) among the best performers at the DREAM6 Molecular Classification of Acute Myeloid Leukemia Challenge. The description of the challenge can be found at http://www.the-dream-project.org/challenges/dream6flowcap2-molecular-classification-acute-myeloid-leukaemia-challenge.

The approach uses relative entropies to evaluate if the distribution of the values of flow cytometry data for a given individual is closer to the overall distribution for AML or for Normal individuals in the space of values $\Gamma$. Explicitly, the relative entropy difference $\Delta S_i = S_{i,AML} - S_{i,Normal} = -\int P_i(\Gamma)\ln[P_{Normal}(\Gamma)/P_{AML}(\Gamma)]d\Gamma$ indicates that the individual looks like an AML patient for positive values and like a Normal subject for negative values.

Implementation:

1. Compute the 4-dimensional histograms $H(\Gamma)_{i,j,k}$ for the values of $\Gamma$ =("FS Log", "SS Log", "FL3 Log", $j$ ) with $j \in$ {"FL1 Log","FL2 Log", "FL4 Log","FL5 Log"} for Tube $k \in$ {1,2,3,4,5,6,7} for all individuals $i$. For each individual there are $7\times 4 = 28$ histograms.
2. Compute $H(\Gamma)_{AML,j,k} = \sum_{i\in AML} H(\Gamma)_{i,j,k}$ and $H(\Gamma)_{Normal,j,k} = \sum_{i\in Normal} H(\Gamma)_{i,j,k}$ as the overall histograms for AML and Normal individuals.
3. Normalize the histograms to obtain the probabilities $P(\Gamma)_{i,j,k}$, $P(\Gamma)_{AML,j,k}$, and $P(\Gamma)_{Normal,j,k}$.
4. Compute the relative entropy differences $\Delta S_{i,j,k} = -\sum_\Gamma P_{i,j,k}(\Gamma)\ln[P_{Normal,j,k}(\Gamma)/P_{AML,j,k}(\Gamma)]$. The total relative entropy difference is defined as $\Delta S_i = \sum_{j,k} \Delta S_{i,j,k}$ .
5. The likelihood that an individual $i$ is AML positive is quantified as $L_i = 1/(1+e^{-\Delta S_i})$.

Code execution: The data for the challenge are available at http://flowrepository.org/id/FR-FCM-ZZYA. The two python files needed to execute the code are listed below. They can also be downloaded from http://www.ehu.es/biologiacomputacional/soft/.

In a directory of a Unix/OSX machine with the files "series10createdist_tot.py", "series10usedist_tot.py", "DREAM6AMLTrainingSet.csv", and the directory "CSV" with the files

"0001.CSV", "0002.CSV"... execute:

```
$ python -O -u series10createdist_tot.py ; python -O -u
series10usedist_tot.py
$ cat DREAM6_AML_Predictions_team21u.txt | sort -n -k 3 | cut -f 1,2 >
DREAM6_AML_Predictions_team21.txt
```

"series10createdist_tot.py" file:

```
import csv
from pylab import *
import numpy


GetNames= lambda y: ['CSV/%04d.CSV' % ((y-1)*8+i) for i in range(1,8)]
ConR= lambda y: [log10(float(y[0])),float(y[1])]+[float(i) for i in y[2:]]
ReadF= lambda y: [ConR(i) for i in [row for row in csv.reader(open(y,'rb'))][1:]]

myranges=[arange(2.01,3.01,0.1)]+[arange(0.01,1.01,0.1) for i in range(6)]
ibin=[[myranges[1],myranges[4],myranges[0],myranges[i]] for i in [2,3,5,6]]
iex=[(1,4,0,i) for i in [2,3,5,6]]
ll=len(ibin)

def doSomething(xx, hh):
  a=array(ReadF(xx))
  H=range(ll)
  edges=range(ll)
  for i in range(ll):
         H[i], edges[i] = histogramdd(a[:,iex[i]], bins=ibin[i])
  return [H[i]+hh[i] for i in range(ll)]

Tset=[row for row in csv.reader(open('DREAM6_AML_TrainingSet.csv','rb'))]

AMLTs=[int(i[0]) for i in Tset if i[1]=='AML']
NormalTs=[int(i[0]) for i in Tset if i[1]=='Normal']


rr=range(7)

H=[[zeros([len(i)-1 for i in ibin[k]]) for k in range(ll) ] for j in rr]
for j in AMLTs[:]:
  for i in rr:
         print j
         H[i]=doSomething(GetNames(j)[i], H[i])
         for k in range(ll):
                numpy.save('hist_a_%d_%d.npy'% (i, k),H[i][k])

H=[[zeros([len(i)-1 for i in ibin[k]]) for k in range(ll) ] for j in rr]
for j in NormalTs[:]:
  for i in rr:
         print j
         H[i]=doSomething(GetNames(j)[i], H[i])
         for k in range(ll):
                numpy.save('hist_n_%d_%d.npy'% (i, k),H[i][k])
```

"series10usedist_tot.py" file:

```
import csv
```

```python
from pylab import *
import numpy

GetNames= lambda y: ['CSV/%04d.CSV' % ((y-1)*8+i) for i in range(1,8)]
ConR= lambda y: [log10(float(y[0])),float(y[1])]+[float(i) for i in y[2:]]
ReadF= lambda y: [ConR(i) for i in [row for row in csv.reader(open(y,'rb'))][1:]]

myranges=[arange(2.01,3.01,0.1)]+[arange(0.01,1.01,0.1) for i in range(6)]
ibin=[[myranges[1],myranges[4],myranges[0],myranges[i]] for i in [2,3,5,6]]
iex=[(1,4,0,i) for i in [2,3,5,6]]
ll=len(ibin)

def doSomething(xx, hh1, hh2, yn=1):
  a=array(ReadF(xx))
  H=range(ll)
  edges=range(ll)
  for i in range(ll):
         H[i], edges[i] = histogramdd(a[:,iex[i]], bins=ibin[i])
  hh1=[hh1[i]-yn*H[i] for i in range(ll)]
  epsi=[1e-10 for i in range(ll)]
  peq1=[(epsi[i]+hh1[i])/sum(epsi[i]+hh1[i]) for i in range(ll)]
  peq2=[(epsi[i]+hh2[i])/sum(epsi[i]+hh2[i]) for i in range(ll)]
  pne=[(epsi[i]+H[i])/sum(epsi[i]+H[i]) for i in range(ll)]
  slocal1=[pne[i]*log(pne[i]/peq1[i]) for i in range(ll)]
  slocal2=[pne[i]*log(pne[i]/peq2[i]) for i in range(ll)]
  return [sum(slocal1[i])-sum(slocal2[i]) for i in range(ll)]


Tset=[row for row in csv.reader(open('DREAM6_AML_TrainingSet.csv','rb'))]

AMLTs=[int(i[0]) for i in Tset if i[1]=='AML']
NormalTs=[int(i[0]) for i in Tset if i[1]=='Normal']


rr=range(7)

Ha=[[numpy.load('hist_a_%d_%d.npy'%(j, i)) for i in range(ll)] for j in rr]
Hn=[[numpy.load('hist_n_%d_%d.npy'%(j, i)) for i in range(ll)] for j in rr]

for j in AMLTs[:]:
  res=zeros(ll)
  for i in rr:
         rest=doSomething(GetNames(j)[i], Ha[i], Hn[i])
         print j, rest
         res+=rest
  print 0, sum(res), j, 9345677


for j in NormalTs[:]:
  res=zeros(ll)
  for i in rr:
         rest=doSomething(GetNames(j)[i], Hn[i], Ha[i])
         print j, [-i for i in rest]
         res+=rest
  print 1, -sum(res), j, 9345677

f=open("./DREAM6_AML_Predictions_team21u.txt",'w')
for j in range(180,360):
  res=zeros(ll)
  for i in rr:
         rest=doSomething(GetNames(j)[i], Ha[i], Hn[i], 0)
         print j, rest
```

```
            res+=rest
  print 2, sum(res), j, 9345677
  f.write("%d\t%f\t%f\n" % (j,1/(1+exp(sum(res))), sum(res)))
  f.flush()
f.close()
```

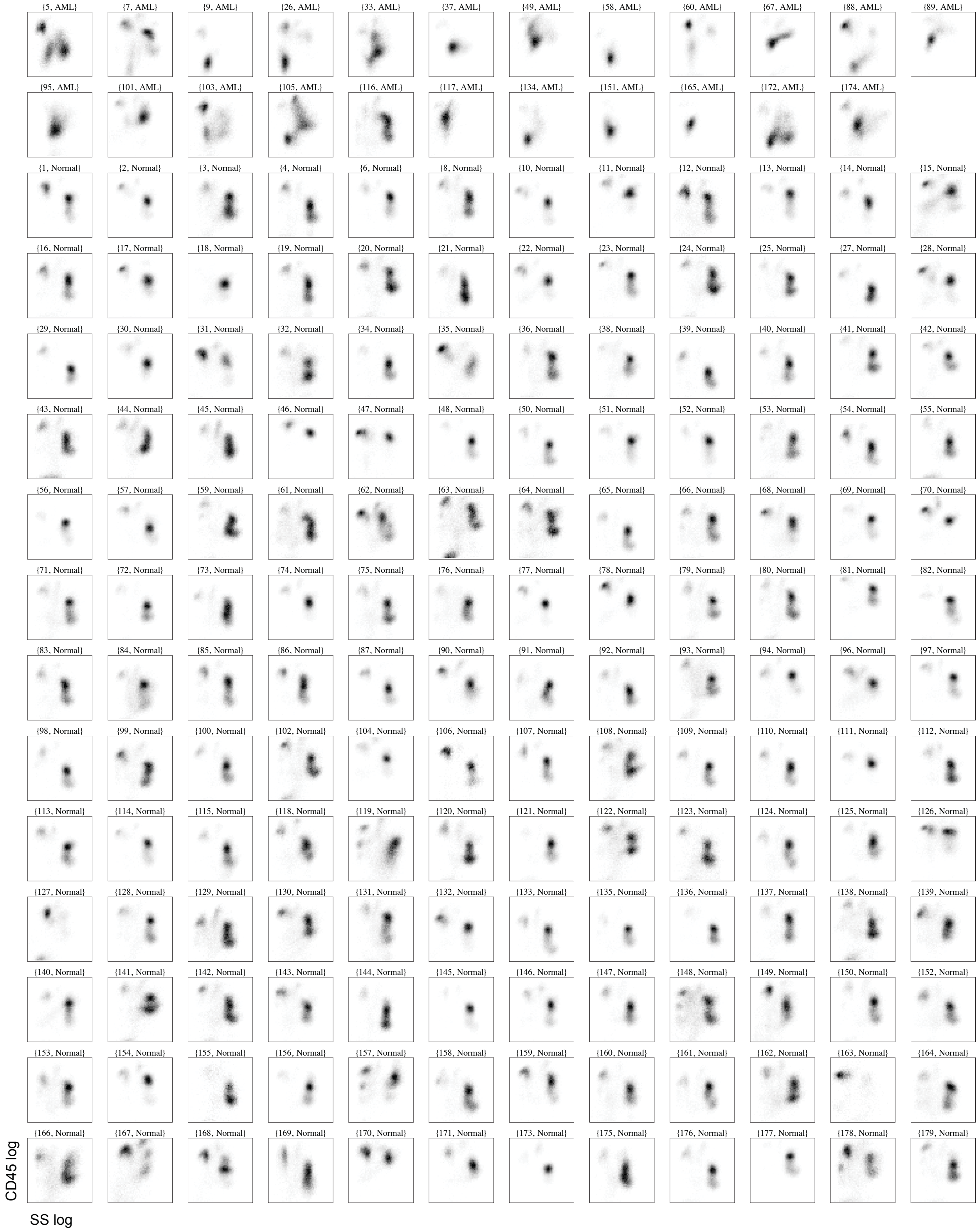

**Supplementary Figure 1: Interpopulation variability of AML and Normal cell populations**. The two-dimensional distribution of cells with given values of the logarithms of the side scatter (SS log) and the marker CD45 (CD45 log) intensities are shown for all cell populations of the training set, with black and white colors representing high and zero densities, respectively. The label on the top of each graph indicates the patient number and the population type.

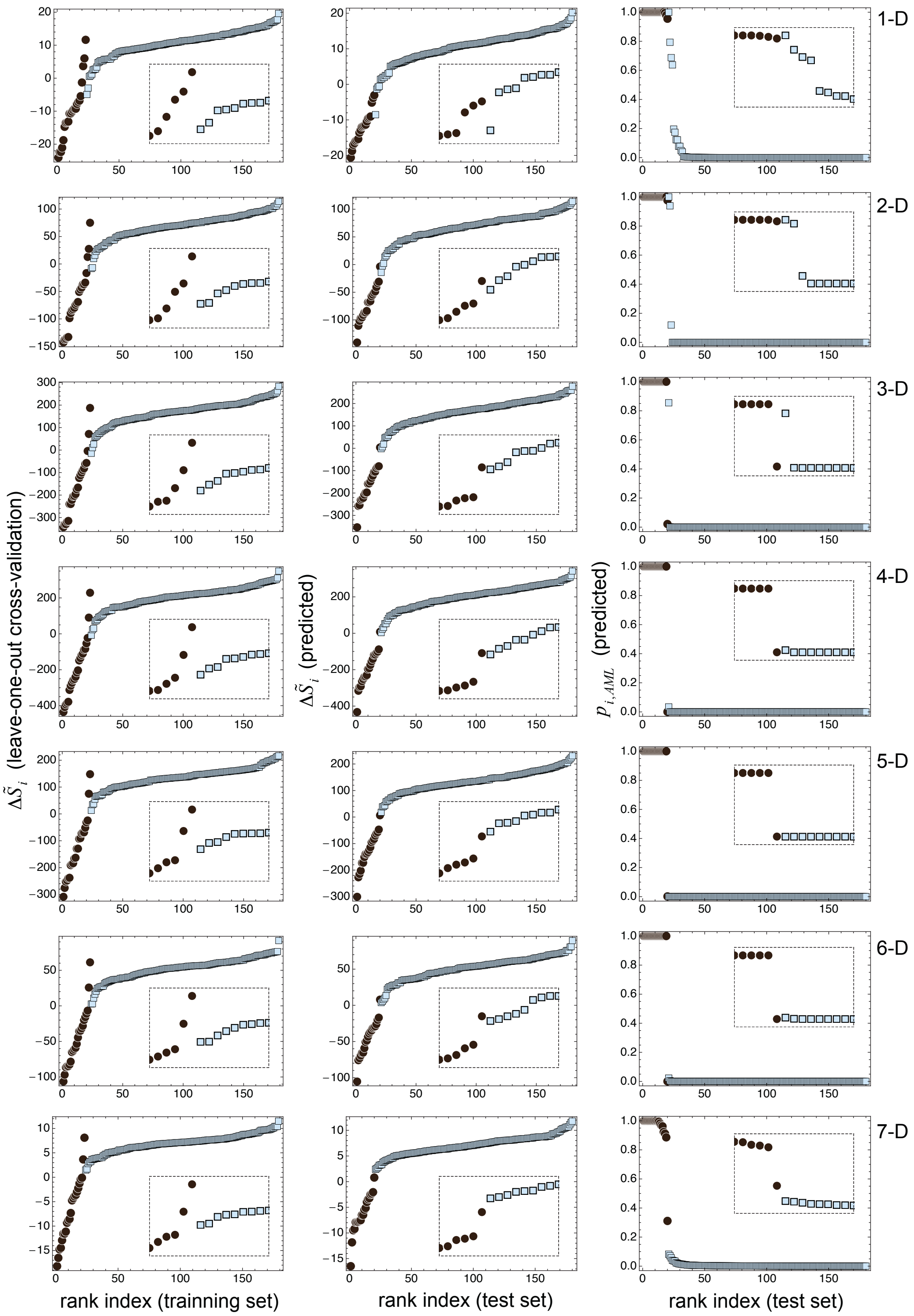

**Supplementary Figure 2: Entropy-based discrimination between AML and Normal individuals**. The total relative entropy difference, using equation (5) with $w_k = 1$, is shown for each individual of the training and test sets from one- to seven-dimensional landscapes. The likelihood that an individual of the test set is AML positive, given by equation (6), is also shown from one- to seven-dimensional landscapes. Circles and squares represent, respectively, the actual AML and Normal state of a patient as clinically assessed by a physician. The insets are magnifications around the AML-Normal boundary.

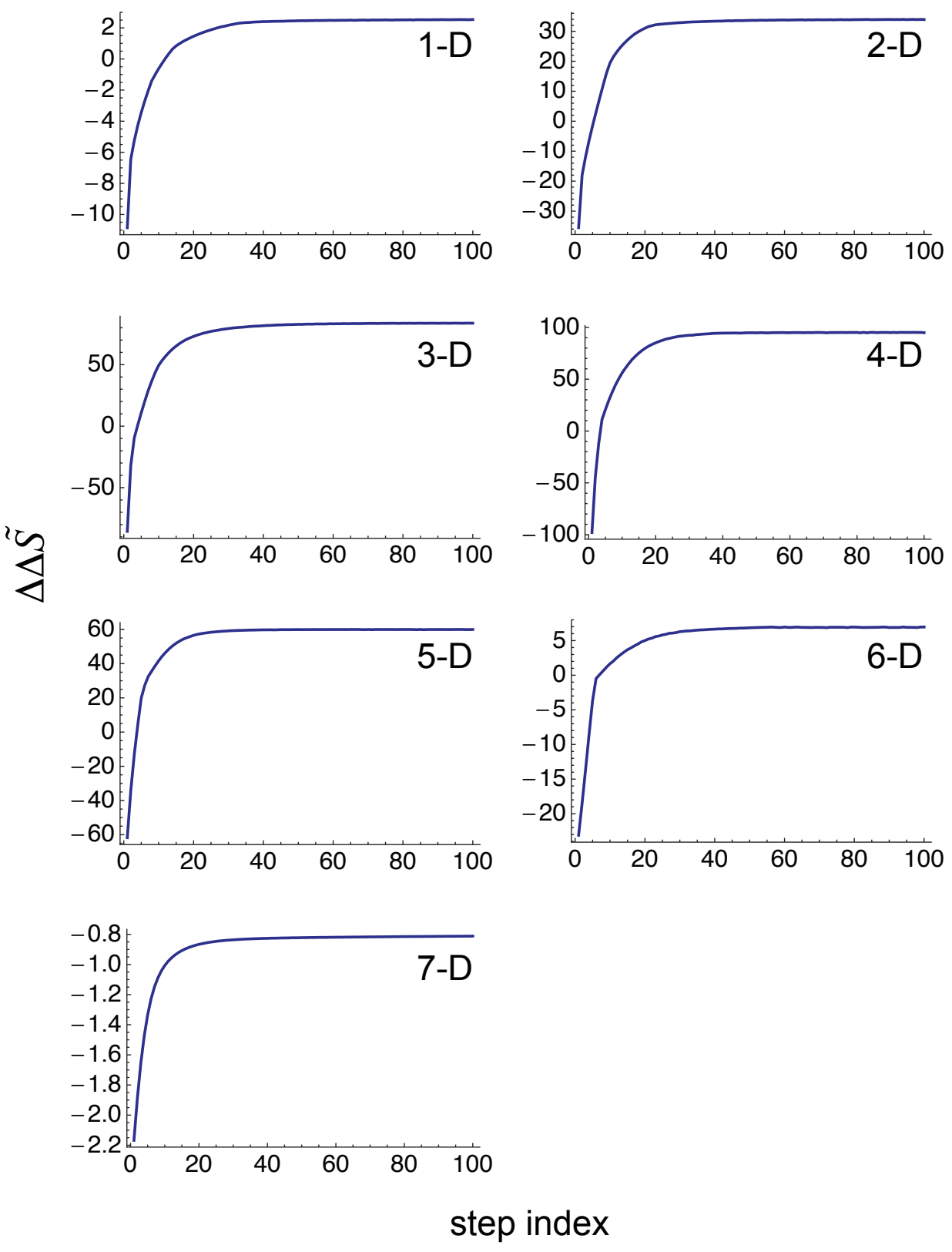

**Supplementary Figure 3: Evolution of the segregation measure.** The value of $\Delta\Delta\tilde{S} = \sum_k w_k \Delta\Delta S^k$ , with $\Delta\Delta S^k = \Delta S^k_{i=N\min} - \Delta S^k_{i=A\max}$ , $w_k(t)$ from equation (7), and $w_k(0) = 1$, is shown for the training set from one- to seven-dimensional landscapes.

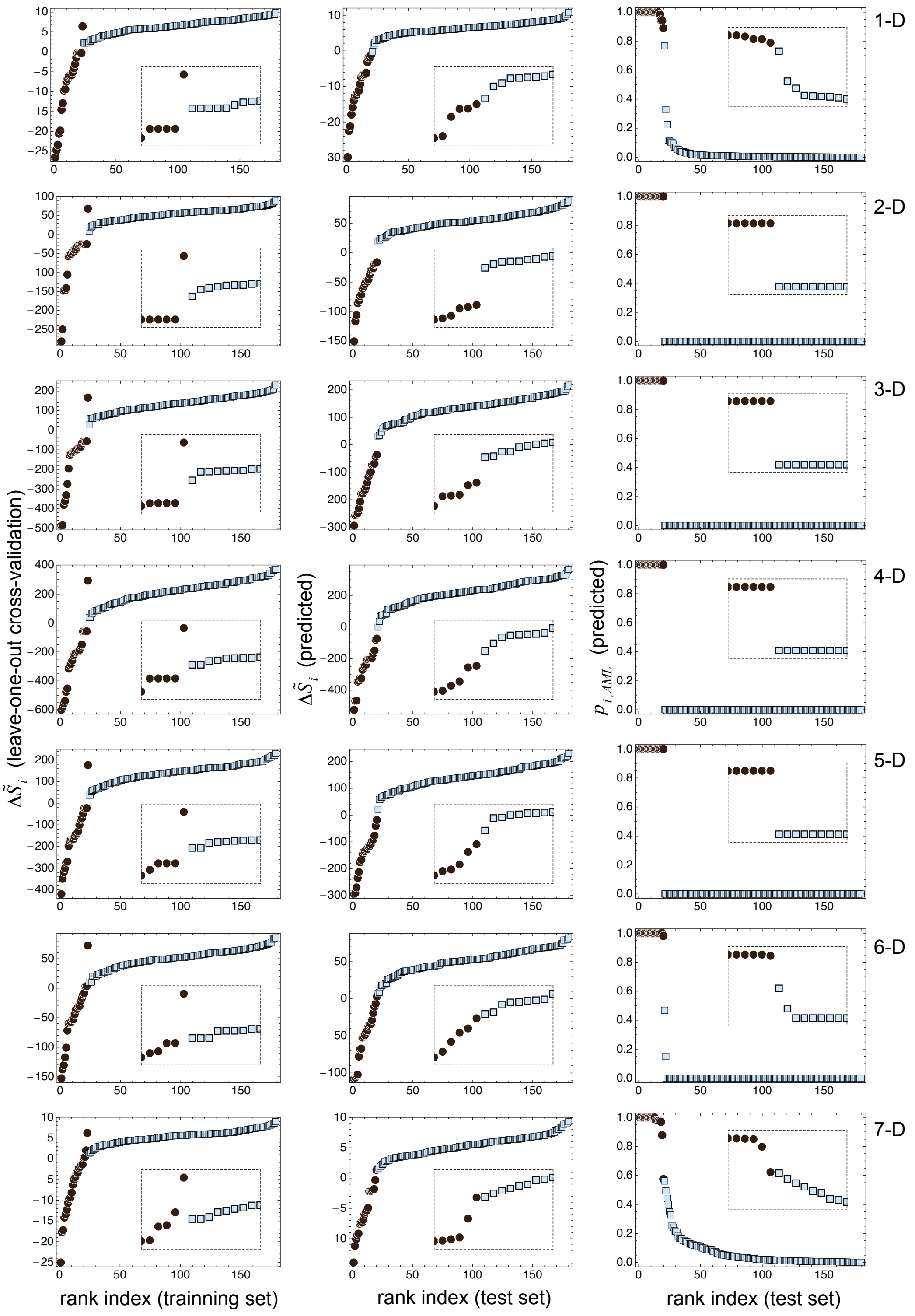

**Supplementary Figure 4: Weighted-entropy enhancement of the discrimination between AML and Normal individuals**. Same situation as in Supplementary Figure 2 but with the weights $w_k$ that maximize the distance $\Delta\Delta\tilde{S}$ of the training set excluding patient #116.